\documentclass[jkps,preprint,fleqn,showpacs,showkeys]{revtex4}
\usepackage{graphicx}
\usepackage{amssymb}
\usepackage{amsmath}
\usepackage{bm}
\def\be{\begin{equation}} \def\ee{\end{equation}} \def\bea{\begin{eqnarray}}
\def\eea{\end{eqnarray}} \def\nnb{\nonumber}
\begin{document}
\setcounter{page}{0}
\title[]{
A renormalization method
for three-boson system with a triboson field
}
\author{Shung-Ichi \surname{Ando}}
\email{sando@sunmoon.ac.kr}
%\thanks{Fax: +82-41-531-1643}
\affiliation{
School of Mechanical and ICT Convergence Engineering,
Sunmoon University,
Asan, Chungnam 31460}

\date[]{Received 16 December 2016}

\begin{abstract}
A renormalization method that introduces an auxiliary field 
to represent a three-body bound state
is studied in a three-boson system
with a triboson field.
A cutoff dependence in the three-boson system emerges
as a limit cycle, and the cyclic singularity is renormalized
by employing two methods: 
a standard method and the auxiliary field method.
For each method, different sets of diagrams
are involved for renormalization,
and thus we numerically study three quantities:
counter term for renormalization,
scattering length of $s$-wave boson-diboson scattering,
and normalized wavefunction.
We confirm that the two methods would lead to the same result.
\end{abstract}



\maketitle

\section{Introduction}

The studies of three-body systems in pionless effective field theory (EFT)
revealed a nontrivial feature of the systems.
That is an appearance of a cyclic singularity, known as a limit cycle,
in a three-boson system~\cite{bhvk-npa99}
and a three-nucleon system in triton channel~\cite{bhvk-npa00}.
A three-body contact interaction, whose order is naively counted as 
that in higher order, is promoted to leading order (LO) for renormalization
of the singularity.
The appearance of the limit cycle is accompanied by emergence of bound
states, known as Efimov states, in the unitary limit~\cite{efimov}.
For recent studies of the three-nucleon systems in triton and $^3$He channels,
one may refer to
Refs.~\cite{ab-jpg10,kh-prc11,vetal-prc14,ketal-jpg16}.
(For general reviews of the pionless EFT, one may refer to
Refs.~\cite{bvk-arnps02,bh-pr06}.)
This feature is also applied to the studies of the various systems,
e.g., halo-nuclei~\cite{h-epjwc16}
and hyper-nuclei~\cite{h-npa02,a-ijmpe16,ayo-prc14,ao-prc14,aro-prc15}.

One of the issues of the three-nucleon systems in pionless EFT
is to establish a rigorous perturbative method to expand an amplitude 
in terms of effective range terms
because the major part of the previous works employed an approximation
so called partially resummed approach~\cite{gbg-npa00}.
A fully perturbative method for a calculation of
$nd$ scattering in pionless EFT was suggested by Vanasse.
However, it is not easy to apply the method to a study involving
a bound state~\cite{v-prc13}.
Recently, the same author suggested a new method to deal with
a bound state perturbatively by introducing a tribaryon field,
which represents the bound state of triton, and the method was
applied to a calculation of charge radius of the triton up to
next-to-next-to-leading order~\cite{v-16}.

This new method introduces an auxiliary field which
represents a bound state of a three-body system.
A dressed three-body propagator is constructed
by using the auxiliary field, and a coupling constant
of the auxiliary field is determined so that
a pole position of the three-body binding energy
is reproduced in the dressed three-body propagator.
An advantage of the new method is that it is not necessary
to numerically fit the coupling constant of the three-body
part by employing Newton's method.
The method has originally been introduced by Hagen et al.,
and they applied it to the study of electric form factor
of two-neutron halo systems~\cite{hhp-epja13}.

In this short report,
we study the renormalization method suggested
by Hagen et al. by employing a simple system,
three neutral scalar boson system of equal masses
with a triboson field.
Using a simple system is beneficial
to a study of the renormalization method itself
since it avoids detailed calculations.
The system exhibits a limit cycle,
thus a sharp momentum cutoff $\Lambda$ is introduced
in an integral equation for the three-boson system,
and a value of a three-body coupling is determined
as functions of $\Lambda$
by employing two methods: a standard method and the auxiliary field method.
In the two methods, as to be discussed in detail later,
sets of diagrams involved in the scattering amplitude are the same.
However, those for renormalization and calculations of a normalized
wavefunction are different.
So it would be worth confirming that the two different methods
lead to an identical result.
We then numerically calculate scattering length of $s$-wave
boson-diboson scattering and normalized wavefunctions,
and we confirm that the two method would lead to the same result.

This work is organized as following.
In Sec. 2, an effective Lagrangian
for a three-boson system with a triboson field
is displayed, and in Sec. 3, two-body and three-body parts
of an equation for the three-body system are
constructed from the Lagrangian for the two renormalization methods.
In Sec. 4, numerical results are obtained, and finally in Sec. 5,
results and discussion of the work are presented.

\section{Lagrangian}

To study the renormalization method for a three-body system,
we consider a simple system, three neutral scalar bosons having
equal masses, which makes a two-body bound state and a three-body bound state.
In addition, we employ standard counting rules in pionless
EFT for two and three-body systems~\cite{bhvk-npa00,bvk-arnps02}.
Here we consider LO contributions only.
Thus we employ a simple Lagrangian for the three-boson system
including diboson and triboson fields
as~\cite{hhp-epja13,brgh-npa03,g-npa04,bs-npa01,ah-prc05}
\bea
{\cal L} &=& \phi^\dagger \left(
i\partial_0 + \frac{1}{2m}\nabla^2
\right) \phi
+ \cdots
\nnb \\ &&
+ d^\dagger \Delta_d d
- \frac12 y_d \left(d^\dagger \phi\phi + \phi^\dagger\phi^\dagger d\right)
+ \cdots
\nnb \\ &&
+ t^\dagger \Delta_t t
- y_t \left(t^\dagger d\phi + d^\dagger\phi^\dagger t\right)
+ \cdots
\,,
\eea
where $\phi$, $d$, and $t$ are boson, diboson, and triboson fields,
respectively, and $m$ is the boson mass.
The dots denote higher order terms which have more derivatives.
Four parameters, $\Delta_d$, $\Delta_t$, $y_d$, and $y_t$,
appear in the Lagrangian at LO. 
$\Delta_d$ and $\Delta_t$ are fixed by using two and three-body
binding energies, $B_2$ and $B_3$,
respectively, whereas $y_d$ and $y_t$ can arbitrarily be chosen.

\section{Amplitudes}

Two-body and three-body parts
of an amplitude for the three-boson system are presented
in the following, and  we discuss a limit cycle appearing
in the three-body part.

\subsection{Two-body part}

\begin{figure}[t]
\includegraphics[width=10cm]{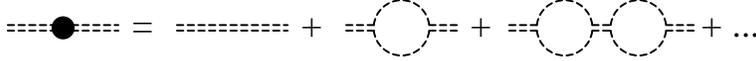}
\caption{
Diagrams for propagator of dressed diboson field.
A double (single) dashed line denotes bare diboson (boson) field.
}
\label{fig;propagator}
\end{figure}

In Fig.~\ref{fig;propagator}, diagrams of a dressed diboson
propagator are depicted where the two-boson bubble diagrams
are summed up to infinite order.
One may have the renormalized dressed diboson propagator as~\cite{ah-prc05}
\bea
D_d(p_0,\vec{p}) &=&
\frac{1}{
\gamma_d
- \sqrt{
-mp_0 + \frac14\vec{p}^2
-i\epsilon
}
}\,,
\label{eq;Dd}
\eea
where $p_0$ and $\vec{p}$ are off-shell energy and three
momentum of the propagation of the diboson state,
and parameters in the propagator have been fixed as
$y_d^2 = \frac{8\pi}{m}$,
$\gamma_d = \sqrt{mB_2} = \Delta_d$
where $\gamma_d$ is the binding momentum of the two-boson
bound state.
The wavefunction normalization factor $Z_d$
of the diboson field is obtained by using the relation,
$Z_d^{-1} = \left.\frac{d D_d^{-1}(E,\vec0)}{dE}\right|_{E=-B_2}$,
as
\bea
Z_d = \frac{2\gamma_d}{m}\,.
\label{eq;Zd}
\eea

\subsection{Three-body part}

We construct the three-body part in two ways.
We refer to a conventional method as
``standard renormalization method'' (SM) and,
to the one suggested by Hagen et al., 
the ``auxiliary field renormalization method'' (AM) in the following.

\subsubsection{Standard renormalization method (SM)}

\begin{figure} [t]
\includegraphics[width=13cm]{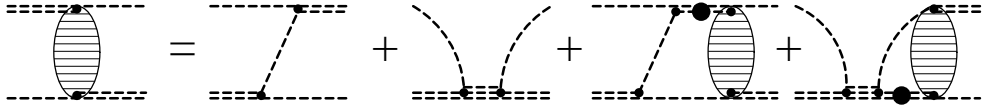}
\caption{
Diagrams for integral equation of $s$-wave boson-diboson scattering.
A shaded blob denotes an elastic scattering amplitude, and
a triple dashed line does propagation of a bare triboson field.
See the caption of Fig.~\ref{fig;propagator} as well.
}
\label{fig;integral_equation}
\end{figure}

In Fig.~\ref{fig;integral_equation},
diagrams of an integral equation for $s$-wave boson-diboson scattering
in terms of a scattering amplitude are depicted.
Thus one has an integral equation from the diagrams in
Fig.~\ref{fig;integral_equation} as~\cite{bhvk-npa99}
\bea
\lefteqn{
t(p,k) =
\frac{4\pi}{pk}
\ln\left(
\frac{p^2 + k^2 + pk -mE}
     {p^2 + k^2 - pk -mE}
\right)
- \frac{y_t^2}{\Delta_t}
}
\nnb \\ &&
- \frac{2}{(2\pi)^2}\int^\Lambda_0 dl l^2\left[
\frac{4\pi}{pl}
\ln\left(
\frac{p^2 + l^2 + pl -mE}
     {p^2 + l^2 - pl -mE}
\right)
- \frac{y_t^2}{\Delta_t}
\right]
\frac{t(l,k)}{
\gamma_d
- \sqrt{
-mE
+\frac34 l^2
}
}\,,
\label{eq;ordinary}
\label{eq;t}
\eea
where $t(p,k)$ is the scattering amplitude of $s$-wave boson-diboson
scattering and $p$ ($k$) is the magnitude of off-shell (on-shell)
relative momentum in final (initial) boson-diboson state
in center of mass frame. A sharp cutoff $\Lambda$ is introduced
for renormalization in the equation.
For the renormalization using the three-body binding energy,
the homogeneous part of the integral equation
in Eq.~(\ref{eq;t}) is numerically solved
by choosing $E=-B_3$,
and the parameter $y_t^2/\Delta_t$
is fitted as a function of $\Lambda$.
The numerical method to solve the equation is described
in Ref.~\cite{ab-jpg10}.

\subsubsection{Auxiliary field renormalization method (AM)}

\begin{figure}[t]
\includegraphics[width=6cm]{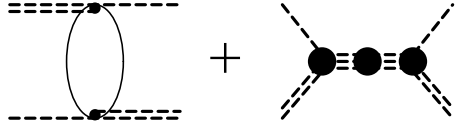}
\caption{
Diagrams for scattering amplitude
for $s$-wave diboson-boson scattering.
}
\label{fig;amplitude}
\end{figure}

\begin{figure}[t]
\includegraphics[width=8cm]{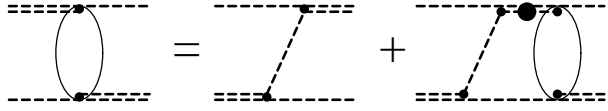}
\caption{
Diagrams for integral equation of scattering amplitude
without triboson propagator.
See the caption of
Fig.~\ref{fig;propagator} as well.
}
\label{fig;a-amplitude}
\end{figure}

For the auxiliary field renormalization method (AM),
Feynman diagrams of the on-shell scattering amplitude $t(k,k)$
are depicted in Fig.~\ref{fig;amplitude}.
The scattering amplitude $t(k,k)$ is represented as~\cite{v-16,hhp-epja13}
\bea
t(k,k) = a(k,k) + b(k,k)\,,
\label{eq;t=a+b}
\eea
where $a(k,k)$ is a scattering amplitude without including
the triboson field.
In Fig.~\ref{fig;a-amplitude},
diagrams of the scattering amplitude $a(k,k)$ are depicted, and
$a(k,k)$ is calculated by solving the integral equation
in Eq.~(\ref{eq;t}) without including the triboson field.
On the other hand,
$b(k,k)$ is a scattering amplitude through the propagation of
a dressed triboson field,
\bea
b(k,k)  = - Y_t(k,E)D_t(E)Y_t(k,E) \,,
\eea
where
$Y_t(k,E)$ is a dressed boson-diboson-triboson vertex function,
and $D_t(E)$ is a dressed triboson propagator.
\begin{figure}[t]
\includegraphics[width=8cm]{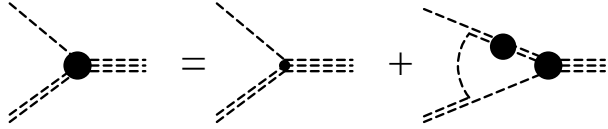}
\caption{
Diagrams for integral equation of three-point
boson-diboson-triboson vertex.
A three-point vertex with a filled circle (a dot)
denotes a dressed (bare) vertex. 
See the caption of
Fig.~\ref{fig;propagator} as well.
}
\label{fig;vertex}
\end{figure}
In Fig.~\ref{fig;vertex}, diagrams of the dressed
boson-diboson-triboson vertex $Y_t(p;E)$ are depicted.
Thus one has an integral equation for $Y_t(p;E)$ as
\bea
Y_t(p,E) &=& y_t
- \frac{2}{\pi}
\int^\Lambda_0 dl l^2
\frac{1}{pl}\ln\left(
\frac{p^2 + l^2 + pl -mE}
     {p^2 + l^2 - pl -mE}
\right)
\frac{Y_t(l,E)}{
\gamma_d
- \sqrt{
-mE + \frac34l^2
}
}\,.
\label{eq;Y}
\eea
\begin{figure}[t]
\includegraphics[width=12cm]{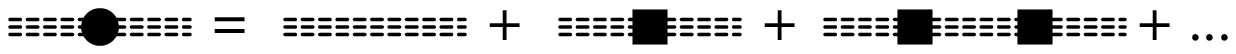}
\caption{
Diagrams for dressed triboson propagator.
A triple dashed line denotes a bare triboson propagation
and a filled box does a self-energy term obtained
in Fig.~\ref{fig;self-energy-triboson}.
}
\label{fig;triboson-propagator}
\end{figure}
In Fig.~\ref{fig;triboson-propagator}, diagrams
for the dressed triboson propagator, which
is obtained by summing a self-energy term up to infinite order,
are depicted.
Thus, from the diagrams, one has the dressed triboson propagator as
\bea
D_t(E) = \frac{1}{\Delta_t - \Sigma_t(E)}\,,
\label{eq;t-propagator}
\eea
where $\Sigma_t(E)$ is the self-energy term whose diagrams are depicted
in Fig.~\ref{fig;self-energy-triboson}.
\begin{figure}[t]
\includegraphics[width=5cm]{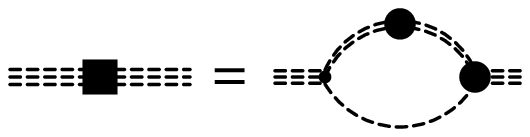}
\caption{
Diagrams for self-energy of the triboson propagation.
See the captions of Figs.~\ref{fig;propagator}
and \ref{fig;vertex} as well.
}
\label{fig;self-energy-triboson}
\end{figure}
Thus we have
\bea
\Sigma_t(E) = \frac{y_t}{2\pi^2}\int^\Lambda_0 dl
\frac{l^2 Y_t(l,E)}{
\gamma_d
-\sqrt{
-mE + \frac34l^2
}
}\,.
\label{eq;Sigma-t}
\eea
The parameter $\Delta_t$ is fixed so as to reproduce the pole structure
of the three-body binding energy at $E=-B_3$ in the dressed triboson
propagator.
We note that the diagrams involving in the scattering state
in Figs.~\ref{fig;integral_equation} and \ref{fig;amplitude} for SM and AM,
respectively, are the same, whereas those for renormalization using
the three-body binding energy are different.
For example, an infinite loop diagram due to
one-boson-exchange interaction is included for renormalization in SM,
whereas this term exists in the amplitude $a(k,k)$
and is excluded for renormalization in AM.

\subsection{Limit cycle of the integral equations}

The integral equations for $t(p,k)$, $a(p,k)$,
and $Y_t(p,E)$
become the same 
in asymptotic limit where $\Lambda\to \infty$, $p,l >> E,\gamma_d,k$,
and thus one has, e.g., for $t(p,k)$ as~\cite{bhvk-npa99,a-ijmpe16}
\bea
t(p) &=& \frac{4}{\sqrt3\pi}\int^\infty_0 \frac{dl}{p}
\ln\left(
\frac{p^2+l^2+pl}{p^2+l^2-pl}
\right) t(l)\,,
\label{eq;asymptotic-ie}
\eea
where the $k$ dependence in $t(p,k)$ is dismissed above.
Then the integral equation becomes scale free,
and that indicates a power behavior of the amplitudes
in the asymptotic limit,
\bea
t(p,k)\,, a(p,k)\,, Y_t(p,E) \propto p^{s-1}\,.
\label{eq;asymptotic-t}
\eea
After performing a Mellin transformation in Eq.~(\ref{eq;asymptotic-ie})
using the relation in Eq.~(\ref{eq;asymptotic-t}),
one has~\cite{bhvk-npa99,j-12,jp-fbs13}
\bea
1 = \frac{8}{\sqrt3s}\frac{\sin\left(
\frac16\pi s\right)}{\cos\left(\frac12\pi s\right)}\,.
\eea
The solution of $s$ for the equation becomes imaginary,
$s=\pm i s_0$ and $s_0 = 1.0064\cdots$.
The imaginary solution indicates the emergence
of a limit cycle. The limit cycle exhibiting in $t(p,k)$ is renormalized
by the three-body counter term in SM,
whereas those in $a(p,k)$ and $Y_t(p,E)$ (as well as $\Sigma_t(E)$)
are not in AM.
They will be sensitive to a value of the cutoff $\Lambda$,
and bound states emerge as Efimov states in them.

Apart from the diagrams involving in the renormalization 
are different in the two methods, 
when the dressed three-body propagator $D_t(E)$
is renormalized in AM so as to reproduce the three-body binding pole in it,
other bound states can emerge as Efimov states
in the other parts, $a(k,k)$ and $Y_t(k,E)$, in the amplitude $t(k,k)$.
In addition, when a normalized wavefunction is derived in AM, the wavefunction
is obtained from the amplitude $b(k,k)$, and the amplitude $a(k,k)$ is excluded
from the derivation of the wavefunction even though bound states are
generated in it.
Those observations are the main concern in the present work,
and we are going to numerically study them in the next section.

\section{Numerical results}

The three-boson system we study in the present work
may not have a corresponding real physical system.
Thus, for numerical study of the renormalization method
in the three-boson system,
we employ values of mass, two and three-body binding energies
of the three-nucleon system in triton channel:
\bea
&&
m = 940~\mbox{\rm MeV}\,,
\ \ \
B_2 = 2.22~\mbox{\rm MeV}\,,
\ \ \
B_3 = 8.48~\mbox{\rm MeV}\,,
\eea
and the binding momentum $\gamma_d$ for the diboson system
is $\gamma_d = \sqrt{mB_2} \simeq 45.7~\mbox{\rm MeV}$.

\subsection{Renormalization at three-body binding energy}

The integral equations in Eqs.~(\ref{eq;t}) and (\ref{eq;Y})
exhibit a limit cycle, and we renormalize the cyclic singularity
at the three-body binding energy, $E=-B_3$,
with a given value of the cutoff $\Lambda$
by employing the two methods, SM and AM.

For the standard renormalization method, SM, we solve
the homogeneous part of the integral equation in Eq.~(\ref{eq;t})
employing a standard expression of the counter term as~\cite{bhvk-npa99}
\bea
\frac{H(\Lambda)}{\Lambda^2}
=
-\frac{y_t^2}{8\pi \Delta_t}
\,.
\label{eq;H}
\eea
For the auxiliary field renormalization method, AM, on the other hand,
the coupling constant $\Delta_t$ is determined by using the pole position
of three-body bound state in the dressed triboson propagator
in Eq.~(\ref{eq;t-propagator}) as~\cite{hhp-epja13}
\bea
\Delta_t - \Sigma_t(-B_3) = 0\,.
\label{eq;BM}
\eea
Using Eq.~(\ref{eq;H}) we have
\bea
H(\Lambda ) = -\frac{\Lambda^2}{8\pi}\frac{y_t^2}{\Sigma_t(-B_3)}\,,
\label{eq;H-BSRM}
\eea
for AM.
We note that
$H(\Lambda)$ for AM does not depend on $y_t^2$
because of $\Sigma_t(-B_3)\propto y_t^2$.
In addition, as mentioned above, the diagrams for renormalization
involving in SM and AM are different.
In SM, one has $2^n$ $n$-loop diagrams with $n\to \infty$
from the homogeneous part of the integral equation in Eq.~(\ref{eq;t}),
and a value of $H(\Lambda)$ is numerically searched by using
Newton's method at the point where the determinant of the matrix vanishes
with $E=-B_3$.
In AM, on the other hand,
one has all possible bubble diagrams in the self-energy term,
$\Sigma_t(E)$ in Eq.~(\ref{eq;Sigma-t}), and a value of $H(\Lambda)$
is calculated using Eq.~(\ref{eq;H-BSRM}).

\begin{figure}[t]
\includegraphics[width=10cm]{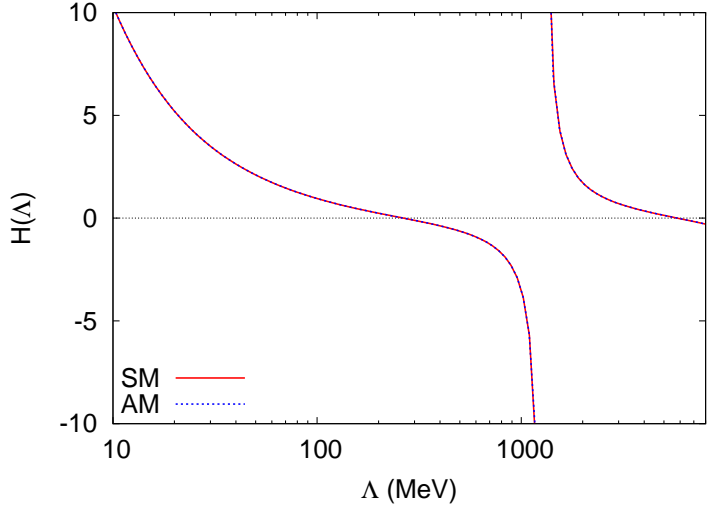}
\caption{
$H(\Lambda)$ as functions of cutoff $\Lambda$
for SM and AM.
}
\label{fig;HvsLam}
\end{figure}

In Fig.~\ref{fig;HvsLam}, we plot curves of $H(\Lambda)$
as functions of $\Lambda$ for SM and AM.
For the both methods, SM and AM,
we reproduce a cyclic pattern
in the counter term $H(\Lambda)$
representing a limit cycle for the three-boson system.
In addition, even though the diagrams involving in the two
renormalization methods are different,
we obtained the same curves of $H(\Lambda)$ for SM and AM.

\begin{figure}[t]
\includegraphics[width=10cm]{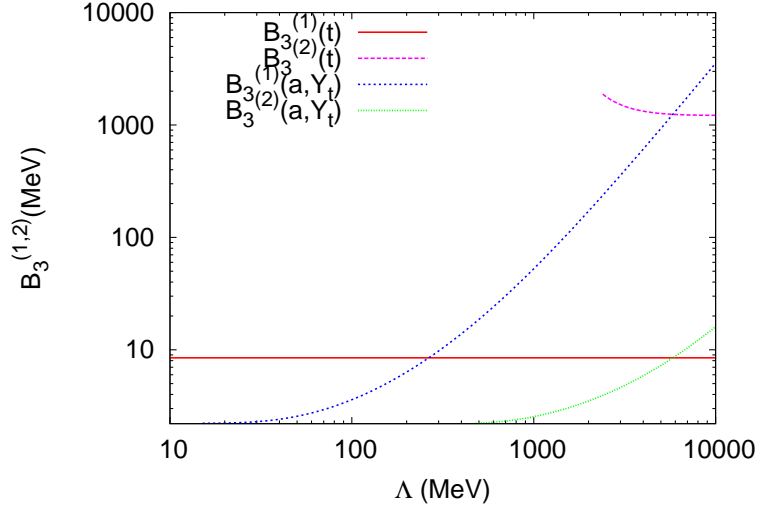}
\caption{First and second binding energies $B_3^{(1)}$ and $B_3^{(2)}$
appearing in amplitudes $t$ and $a$ and dressed vertex function $Y_t$
as functions of $\Lambda$. See the text for details.
}
\label{fig;BvsLam}
\end{figure}

In Fig.~\ref{fig;BvsLam}, we plot curves of first and second
binding energies, $B_3^{(1)}$ and $B_3^{(2)}$,
which appear due to the limit cycle
in the amplitudes $t$ and $a$ and
the dressed vertex function $Y_t$,
as functions of $\Lambda$.
Because the first binding energy $B_3^{(1)}$ appearing in the amplitude $t$
is used as input for renormalization in SM,
we have a horizontal line which corresponds to $B_3^{(1)} = B_3=8.48$~MeV.
A curve of second binding energy, $B_3^{(2)}\sim 1.2$~GeV,
in the amplitude $t$ appears almost flat and
starts around $\Lambda\simeq 1300$~MeV.
One can expect that such a deep binding energy will not affect physics
at low energies.
Meanwhile, one may be concerned that the system
can decay into the deeply bound state once it is formed.
Thus one may choose an upper limit of the cutoff value less than
$\Lambda \simeq 1.3$~GeV for the present system.

In AM, the position of the binding energy $B_3$ is reproduced in
the propagator $D_t(E)$ due to the renormalization,
whereas bound states are created
in the other parts, $a$ and $Y_t$, of the amplitude $t(k,k)$
due to the limit cycle.
First binding energy $B_3^{(1)}$
in $a$ and $Y_t$ starts
appearing around $\Lambda \simeq 15$~MeV.
The binding energy $B_3^{(1)}$ at the starting point is
just above the two-body binding energy, $B_3^{(1)}\simeq B_2=2.22$~MeV,
and it increases as the cutoff value increases.
Second binding energy $B_3^{(2)}$ in $a$ and $Y_t$ starts appearing
around $\Lambda\simeq 480$~MeV
and similarly behaves to the first binding energy $B_3^{(1)}$.
At the point where the second binding energy appears
the first binding energy $B_3^{(1)}$ becomes $B_3^{(1)}\simeq 17.6$~MeV.
Thus $a$ and $Y_t$ contain the small binding energies at the wide range
of the cutoff value, $\Lambda = 15$-10000~MeV in the figure,
and one may expect that those quantities are quite sensitive
to the cutoff $\Lambda$.

\subsection{Scattering length of $s$-wave boson-diboson scattering}

Scattering length $a_3$ of $s$-wave boson-diboson scattering is calculated
by using a formula~\cite{a-fbs14},
\bea
a_3 = - \frac{m}{3\pi}T(0)\,,
\eea
where the scattering matrix $T(E)$  is given as
$T(E) = Z_d t(k,k)$,
with $E=\frac{3}{4m}k^2-B_2$, and the wavefunction normalization
factor $Z_d$ of the diboson field has been presented in Eq.~(\ref{eq;Zd}).
The on-shell amplitude $t(k,k)$ is calculated by solving the
integral equation in Eq.~(\ref{eq;t}) for SM,
and that is given in Eq.~(\ref{eq;t=a+b}) for AM.

\begin{figure}[t]
\includegraphics[width=10cm]{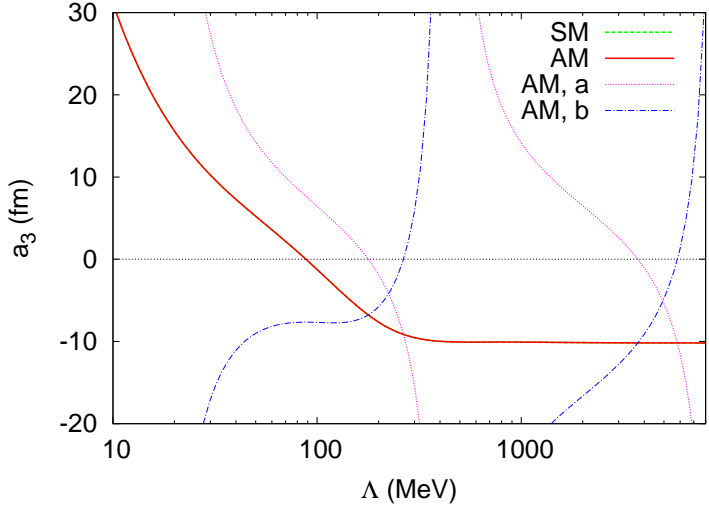}
\caption{
Scattering length $a_3$~(fm) for $s$-wave diboson-boson scattering
as functions of cutoff $\Lambda$~(MeV) for each of renormalization
schemes, SM and AM.
Curves of each contribution from $a(k,k)$ and $b(k,k)$ amplitudes
in Eq.~(\ref{eq;t=a+b}) for AM are also plotted.
}
\label{fig;avsLam}
\end{figure}
In Fig.~\ref{fig;avsLam},
curves of the scattering length $a_3$ 
for the renormalization methods SM and AM
are plotted as functions of the cutoff $\Lambda$. 
Curves of $a_3$ from each component of the amplitudes
$a(k,k)$ and $b(k,k)$ in Eq.~(\ref{eq;t=a+b}) for AM are also included
in the figure.
We find the same curve of $a_3$ for SM and AM.
It is natural because the diagrams of the scattering amplitude
as well as the values of $H(\Lambda)$ are the same for SM and AM.
As seen in the figure, at small cutoff values
$a_3$ are positive.
As the value of the cutoff $\Lambda$ increases, the value of $a_3$
decreases. And when the value of $\Lambda$ becomes larger than
about $\Lambda=300$~MeV, $a_3$ converges to $a_3\simeq -10$~fm.
The behavior of $a_3$ at the small cutoff values implies an artificial
effect that an important part of the momentum flow in the loops
is missing at such small cutoff values~\cite{a-fbs14}.
For AM, a contribution from each of the components $a(k,k)$ and $b(k,k)$
in Eq.~(\ref{eq;t=a+b}), as discussed in the previous subsection,
is sensitive to the cutoff $\Lambda$ and exhibits the limit cycle.
But after they are added together,
one can obtain the cutoff-independent result
at $\Lambda > 300$~MeV.

\subsection{Normalized wavefunctions}

In SM,
a wavefunction for relative boson-diboson part of the three-boson
bound state is calculated by solving the homogeneous part of the
integral equation in Eq.~(\ref{eq;t}).
Because the kernel of the integral equation depends on the energy,
a normalization condition of the wavefunction becomes nontrivial~\cite{lmt-pr65}.
For a wavefunction $|\Psi\rangle$, which
satisfies an equation $|\Psi\rangle = DK|\Psi\rangle$,
one has a normalization condition of the wavefunction as~\cite{kh-prc11}
\bea
\langle \psi |D \frac{d}{dE}\left(
D^{-1} - K
\right) D|\psi\rangle = 1\,,
\eea
where $|\Psi\rangle = D|\psi\rangle$.
In our case, the operators $D$ and $K$ are realized as
$D\to D_d$ in Eq.~(\ref{eq;Dd}), and
\bea
K \to
K(p,l;E) = \frac{4\pi}{pl}\ln\left(
\frac{p^2+l^2+pl -mE}
     {p^2+l^2-pl -mE}
\right)\,,
\eea
and thus we have a normalization condition of the wavefunction as
\bea
&& \frac{m}{4\pi^2}\int^\Lambda_0 dll^2 \left(
\frac{\phi_S(l)}{\gamma_d - \sqrt{mB_3+\frac34l^2}}
\right)^2
\frac{1}{\sqrt{mB_3 + \frac34l^2}}
\nnb \\ &&
- \frac{2m}{\pi^3}
\int^\Lambda_0 dl'l'^2
\int^\Lambda_0 dll^2
\frac{\phi_S(l')}{\gamma_d - \sqrt{mB_3+\frac34l'^2}}
\frac{1}{
(l'^2 + l^2 + mB_3)^2 -l'^2l^2
}
\frac{\phi_S(l)}{\gamma_d - \sqrt{mB_3+\frac34l^2}}
\nnb \\ && = 1\,,
\eea
where $\phi_S(p)=\left.t(p,k)\right|_{E=-B_3}$
is a wavefunction of the bound state obtained in SM.

A normalized wavefunction $\phi_A$ is obtained
from the dressed vertex function $Y_t(p,E)$ for AM as~\cite{hhp-epja13}
\bea
\phi_{A}(p) = -\sqrt{Z_t}Y_t(p,-B_3)\,,
\label{eq;WF}
\eea
where we have included an overall minus sign
in the expression above
so as to obtain a positive value
of the wavefunction at $p=0$.
$Z_t$ is the wavefunction normalization factor of the dressed
triboson field and is obtained from the propagator
$D_t(E)$ in Eq.~(\ref{eq;t-propagator}).
Thus one has
\bea
Z_t = -  \frac{1}{\Sigma_t'(-B_3)}\,,
\eea
where
$\Sigma'_t(E) = \frac{d}{dE}\Sigma_t(E)$.
We note that the wavefunction $\phi_A$ does not depend on $\Delta_t$,
which is used for normalization in Eq.~(\ref{eq;BM}).
In addition, $\phi_A$ does not depend on $y_t$ either
because the coupling $y_t$ is cancelled between those in the vertex $Y_t(p,E)$ 
and the normalization factor $\sqrt{Z_t}$.
Thus one cannot make the wavefunction $\phi_A(p)$ cutoff independence
by adjusting the parameters, $y_t$ and $\Delta_t$.
Furthermore, 
a part of the amplitude, $a(k,k)$, which exhibits a limit cycle
and generates bound states in it, is not included in the
calculation of $\phi_A(p)$.
It might be interesting to examine if
the wavefunction $\phi_A$ can be cutoff-independence
(in other words, if the limit cycle in $Y_t(p,-B_3)$ can be cancelled with
that in the wavefunction normalization factor $\sqrt{Z_t}$)
along with
whether $\phi_A(p)$ is the identical to $\phi_S(p)$ or not.

\begin{figure}[t]
\includegraphics[width=10cm]{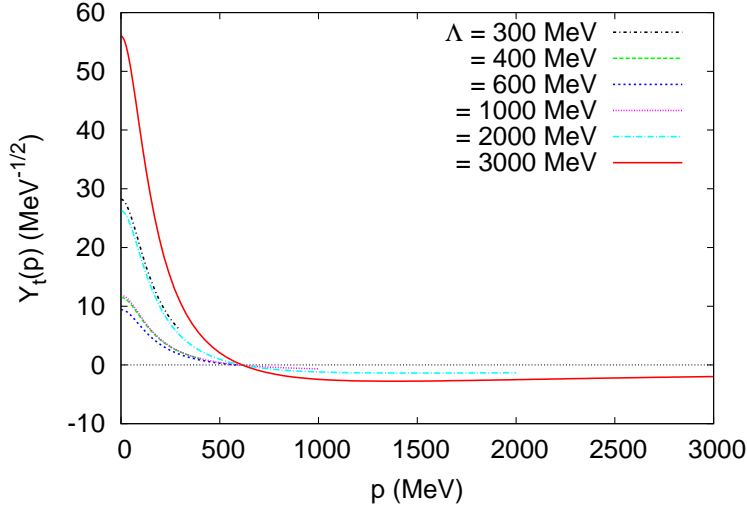}
\caption{
Dressed vertex function $Y_t(p,-B_3)$ as functions of $p$
with various cutoff values $\Lambda=300$ to 3000~MeV in AM
with $y_t = 1$~MeV$^{-1/2}$.
}
\label{fig;Yts}
\end{figure}

\begin{figure}[t]
\includegraphics[width=10cm]{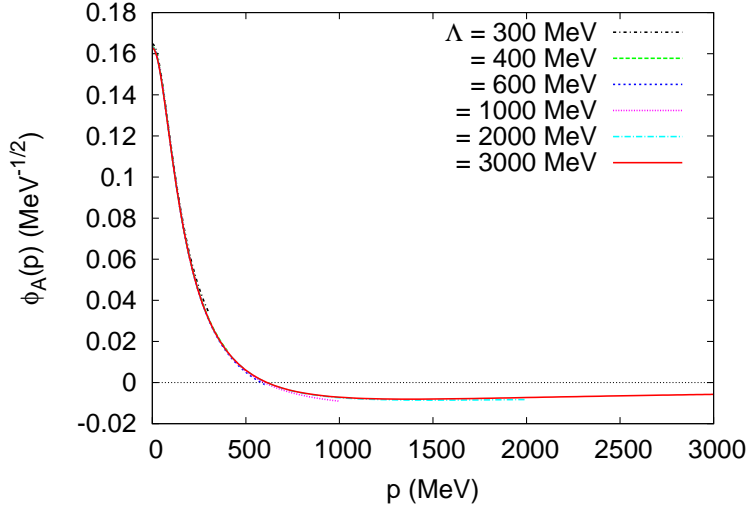}
\caption{
Normalized wavefunction $\phi_A(p)$ as functions of $p$
with various cutoff values $\Lambda=300$ to 3000~MeV for AM.
}
\label{fig;WFs}
\end{figure}

In Figs.~\ref{fig;Yts} and \ref{fig;WFs}, we plot
curves of the dressed vertex function $Y_t(p,-B_3)$ and
those of normalized wavefunctions $\phi_A(p)$, respectively,
as functions of relative momentum $p$ using various cutoff values
from 300 to 3000~MeV for AM where we have used
$y_t=1$~MeV$^{-1/2}$ for $Y_t(p,-B_3)$.
As seen in Fig.~\ref{fig;Yts}, overall factors
of the curves of $Y_t(p,-B_3)$ show a cutoff-dependence,
whereas one can see in Fig.~\ref{fig;WFs}
that the cutoff dependence in the overall factor of $Y_t(p,-B_3)$
disappears in the normalized wavefunction $\phi_A(p)$.
On the other hand, the normalized wavefunction for SM, $\phi_S(p)$,
is indeed cutoff-independent because of the renormalization. 
We find the same $p$-dependence of the normalized wavefunctions
for SM and AM and a difference in the overall factors between them
(we do not show a figure for $\phi_S(p)$).
The factor difference is about 2.36,
and we have $\phi_S(p) \simeq 2.36\, \phi_A(p)$.

\section{Discussion and conclusions}

In the present work, we have studied the auxiliary field renormalization method
suggested by Hagen et al. employing a three-boson system with
a triboson field.
We numerically calculated
the coupling of the three-body contact interaction for renormalization,
the scattering length $a_3$ of $s$-wave boson-diboson scattering, and the
normalized wavefunctions as functions of the cutoff $\Lambda$
employing the two renormalization methods, SM and AM.
We confirm that the three-body system exhibits a limit cycle,
and the cyclic singularity can be renormalized by using the both methods.
Though the different diagrams are involved for renormalization
in those two methods, we obtained the identical result
for the renormalized coupling constant of the three-body contact interaction
as well as the scattering length $a_3$.
In addition, we find that the normalized wavefunctions turned out to be
cutoff independent, however, we obtained the different overall factors
of the normalized wavefunctions in the two methods.

It may be interesting to point out that even though the detailed diagrams
involving in the two renormalization methods and the calculation methods
are different, the results
of the renormalization of the coupling constants turned out to be the same.
Thus, because the diagrams for the scattering amplitude
and the renormalized coupling $H(\Lambda)$ are the same for SM and AM,
it is natural to obtain the same scattering length $a_3$, in which the
limit cycle is renormalized, for the both methods.
On the other hand, the parts of the
scattering lengths $a_3$, $a(0,0)$ and $b(0,0)$, for AM are indeed
sensitive to the cutoff and exhibit the limit cycle.

For the normalized wavefunction $\phi_A(p)$, 
it is also interesting to point out
that it is not necessary to renormalize and fix the parameter $\Delta_t$
in the dressed triboson propagator $D_t(E)$. 
As mentioned above, the two normalized wavefunctions,
$\phi_S(p)$ and $\phi_A(p)$, are obtained from the different diagrams
and the different calculation methods,
while $\phi_S(p)$ and $\phi_A(p)$ turned out to be the same function of $p$
except for the overall factors.
Nevertheless, one would have a same result of a physical observable,
e.g., electric form factor, in the both renormalization methods
after normalizing the wavefunctions by using an available conservative
quantity such as a baryon number or an electric charge.

\begin{acknowledgments}
The author would like to thank M. Birse and J. Vanasse for
discussion and communications.
This work was supported by
the Basic Science Research Program through the National Research
Foundation of Korea funded by the Ministry of Education of Korea
(Grant No. NRF-2016R1D1A1B03930122)
and in part by
the National Research Foundation of Korea (NRF)
grant funded by the Korean government
(Grant No. NRF-2016K1A3A7A09005580).
\end{acknowledgments}

\end{document}